\documentclass[journal,10pt]{IEEEtran}
\usepackage{cite}

\usepackage{graphicx}
\usepackage{url}
\usepackage{times}
\usepackage{amsmath}   
\usepackage{amsfonts,amsthm,amssymb}
\usepackage{mathtools}
\usepackage{nicefrac}
\usepackage[none]{hyphenat}
\usepackage{fontenc}
\usepackage{color}
\usepackage{hyperref}
\usepackage[caption=false,font=footnotesize]{subfig}
\usepackage{url}
\usepackage{balance}
\usepackage{etoolbox}
\usepackage{verbatim} 
\usepackage{float}
\usepackage{placeins}
\DeclareMathOperator{\y}{\mathbf{\hat{y}}}
\DeclareMathOperator{\yb}{\mathbf{y}}
\DeclareMathOperator{\xb}{\mathbf{x}}
\usepackage[dvipsnames]{xcolor}
\usepackage{threeparttable,booktabs}
\usepackage[switch]{lineno}
\begin{document}

    \title{Circularly Polarized Fabry-P\'{e}rot Cavity Sensing Antenna Design using Generative Model}

\author{Kainat~Yasmeen,~\IEEEmembership{Student Member~IEEE}, Kumar~Vijay~Mishra,~\IEEEmembership{Senior Member~IEEE},  A.~V.~Subramanyam,~\IEEEmembership{ Member~IEEE}, and Shobha~Sundar~Ram,~\IEEEmembership{Senior Member~IEEE} 
\thanks{K. Y,  A. V. S., and S. S. R. are with the Indraprastha Institute of Information Technology Delhi, New Delhi 110020 India. E-mail: \{kainaty, subramanyam, shobha\}@iiitd.ac.in.}
\thanks{K. V. M. is with the United States DEVCOM Army Research Laboratory, Adelphi, MD 20783 USA. e-mail: kvm@ieee.org.}
}

\maketitle

\begin{abstract}
We consider the problem of designing a circularly polarized Fabry-P\'{e}rot cavity (FPC) antenna for S-band sensing applications such as satellite navigation and communication. The spatial distribution of peripheral roughness of the unit cell of FPC's partially reflecting surface 
serves as an important design optimization criterion. However, the evaluation of each candidate design using a full-wave solver is computationally expensive. To this end, we propose a deep generative adversarial network (GAN) for realizing a surrogate model that is trained with input-output pairs of antenna designs and their corresponding patterns. Using the GAN framework, we quickly evaluate the characteristics of a large volume of candidate designs and choose the antenna design with an axial ratio of 0.4 dB, a gain of 7.5 dB, and a bandwidth of 269 MHz. 
\end{abstract}

\section{Introduction}
\label{sec:Introduction}
Highly directive radiation from a finite aperture is desired in many electromagnetics applications such as radar, communications, sensing, and imaging \cite{casares2006high,dash2018circularly,rana2017microstrip}. 
Wireless sensor networks for the Internet of Things, vehicle-to-everything (V2X) communications, satellite positioning, and communicating devices require circular polarization (CP) \cite{8568595, zhong2016compact,khidre2012circular}. In these applications, CP offers the advantages of robustness to atmospheric anomalies and rain as well as reduced polarization mismatch loss between a linearly polarized (LP) antenna on a mobile unit with a circularly polarized antenna on a fixed base station. In this context, Fabry P\'{e}rot cavity (FPC) antennas have captured significant research interest because they offer a high gain with a small aperture while retaining a simple feed structure for ease of fabrication and conformal deployment \cite{orr2013design}. The FPC consists of a dielectric cavity encased by a metal ground plane on one side and a partially reflecting surface (PRS) on the other. Excitation from a primary radiator within the cavity undergoes multiple reflections between the PRS and ground plane before emanating from the antenna, resulting in an enhanced gain.

The primary FPC radiators have been conventionally LP microstrip patches or slot antennas. The design of CP FPC antennas is relatively very challenging. 
A common way to achieve CP FPC is to employ a CP primary radiator, such as a feed point at a particular position on an asymmetric and rectangular patch \cite{qing2010compact, sharma1983analysis}. However, this technique suffers from low fabrication error tolerances and narrow bandwidths. Simple patches with dual-orthogonal feed structures have also been proposed \cite{kim2019compact} but a complicated feeding mechanism negates the purpose of realizing a simple single-feed high gain FPC antenna \cite{carver1981microstrip}. Alternatively, an LP primary radiator is used, and the polarization is subsequently changed to CP by the PRS \cite{zhu2013linear,liu2015compact}. 

In \cite{orr2013design}, the PRS was designed with a periodic array of unit cells where each unit cell consists of a rectangular loop with a diagonal. The resulting axial ratio (AR) of the structure was fairly low, but the antenna system was narrowband and of low gain. Our preliminary work in \cite{jain2020circularly} showed that further enhancements in the bandwidth and gain could be realized by incorporating peripheral roughness in the form of bricks along the metal edges of the unit cell. However, there are considerable degrees of freedom (DoFs) in the number and distribution of the bricks along the periphery. This design is not optimal because the evaluation of each candidate design, through either electromagnetic (EM) simulations or measurements, is slow and laborious. In this paper, we propose accelerating the design optimization process by replacing the time-consuming EM simulations with rapid neural network-based surrogate models.

In general, the antenna design often involves the optimization of several complicated and irregular geometry parameters in order to meet multiple objectives pertaining to the resonant frequency, gain, polarization, bandwidth, and size constraints. The procedure involves two time-consuming steps before the optimal design is realized: The first step is the synthesis of multiple candidate designs for evaluation. Traditionally, antenna parameters were optimized through trial and error. Later works have employed evolutionary algorithms, such as genetic algorithms and particle swarm optimization, wherein geometry parameters were optimized through an iterative synthesis of antenna designs constrained by a fitness function on the antenna characteristics \cite{jin2005parallel}.  
In the second step, the antenna characteristics for each design are simulated using computationally-expensive full-wave EM solvers involving finite difference time domain techniques, finite element methods, or method of moments.

Several recent studies have shown to reduce the computational workload during the evaluation process through machine learning (ML) techniques, including artificial neural networks \cite{sedaghat2022compressed}, support vector regression \cite{prado2019wideband}, and deep learning (DL) networks \cite{hodge2019rf,hodge2019joint,liaqat2021hybrid,xiao2018deep}. These methods map the non-linear relationship between the geometric parameters and antenna characteristics using data from EM solvers. Once trained, the design process is significantly accelerated, with the surrogate model replacing the EM solver for rapidly generating the antenna behaviors for any given set of geometry parameters. In this context, deep generative adversarial networks (GANs) have emerged as a preferred DL technique to solve a wide variety of EM problems \cite{alnujaim2021synthesis,liu2018generative,hodge2019multi,liu2021generative, vijayamohanan2020antenna,ye2020inhomogeneous,zhang2022image}. GANs have been shown to be effective for augmentation, classification, and regression problems \cite{hodge2019joint,ye2020inhomogeneous}. As a result, they have been successfully exploited for reconstructing microwave imaging profiles and metasurface design \cite{liu2018generative,ma2021deep,ma2020learning}. Some prior studies employ GAN to obtain antenna parameters from specified antenna characteristics that include basic geometric patterns (circles, pentagons, and hexagons) \cite{liu2021generative}.
A GAN works as a zero-sum game between two deep networks: generator and critic. Its objective is to \emph{implicitly} learn the probabilistic distribution of a set of training samples and, subsequently, create samples of the distribution during the prediction stage \cite{goodfellow2014generative}. The generator produces samples of a distribution from the training data while the critic assesses the samples and decides if they are real or fake (produced by the generator). The primary advantage of using the GAN framework with two competing neural networks is that the GAN is semi-supervised 
and requires smaller, less diverse training data set 
\cite{goodfellow2014generative}. 

In this work, we propose to accelerate the CP FPC parameter optimization through a GAN-aided design procedure. The generator serves as a surrogate model for producing samples of antenna characteristics using training antenna patterns obtained from an EM solver. The inputs to the generator are the unit cell geometric parameters. Once the surrogate model is trained, we use it to simultaneously evaluate the antenna characteristics of several hundred candidate designs, eventually facilitating the choice of the optimal design with bandwidth, AR, and gain as 269 MHz, 0.4 dB, and 7.5 dBi, respectively. We demonstrate the feasibility of our proposed method through validation via full-wave simulations, fabrication of optimized antenna, and hardware measurements. 
\section{FPC Structure}
\label{sec:sysmod} 
In FPC, the primary source of excitation is introduced within the cavity. 
The height of the cavity is carefully chosen such that the multiple reflections within the cavity are in phase with each other when they emanate from the antenna, thereby enhancing the gain of the primary radiator. The polarization of the resulting radiation is determined by either the polarization of the primary radiating source or by the unit cell in the PRS.

\begin{figure}[t]
\centering
\includegraphics[width=3.5in,height=2.5in]{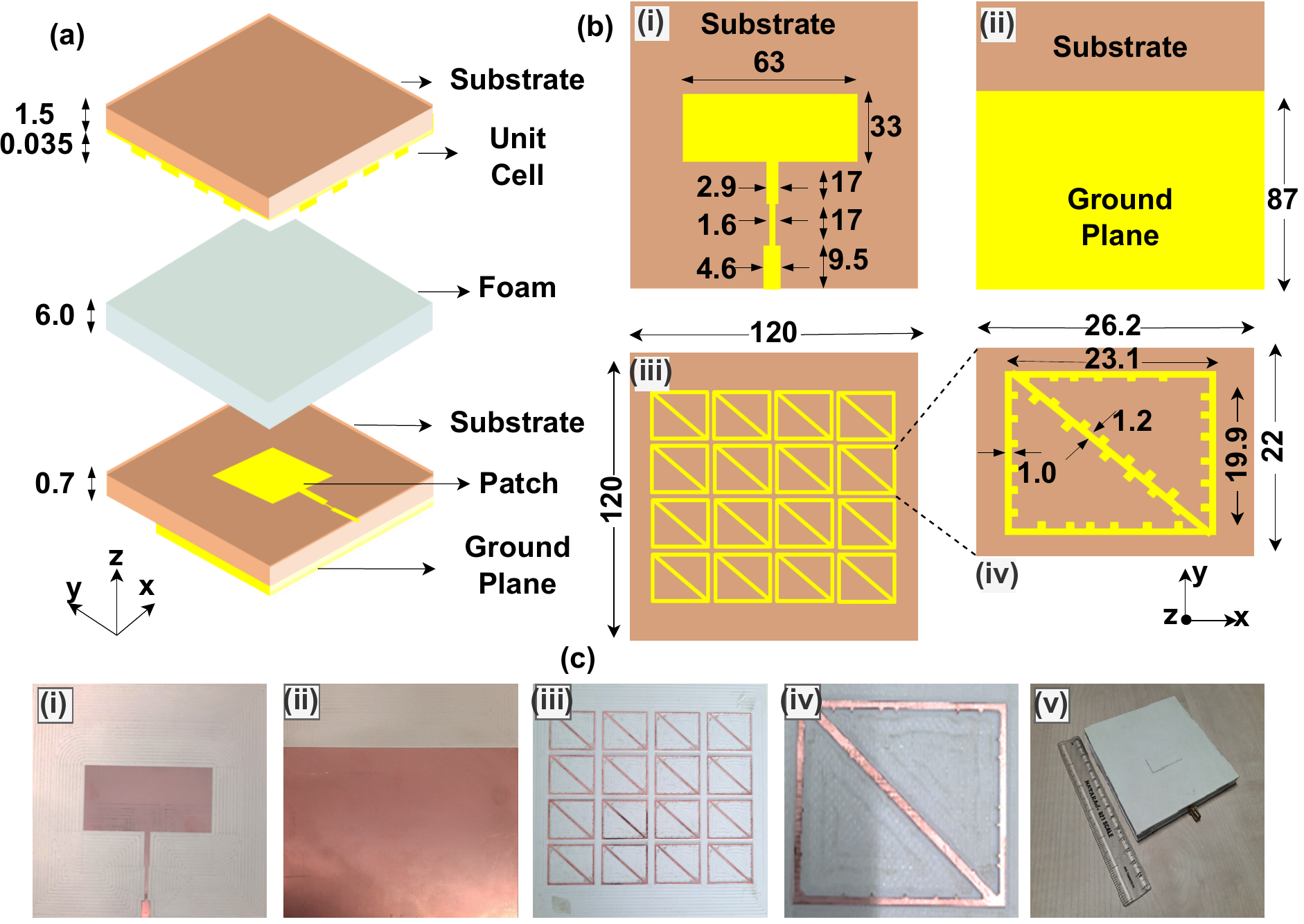}
\vspace{-7mm}
\caption{(a) Three-dimensional (3-D) view of CP FPC antenna. (b) Top-view of (i) patch with a three-stage quarter-wave transformer, (ii) partial reflecting ground plane, (iii) partially reflecting superstrate with $4\times 4$ array of uniform unit cells with the diagonal, and (iv) unit cell with peripheral roughness. (c) As in (b), but for our fabricated FPC antenna with optimized design.}
\label{fig:Antenna}
\end{figure}

Consider a basic FPC structure (Fig.~\ref{fig:Antenna}a), where a single feed patch antenna - the primary radiator resonating at $2.4$ GHz - is mounted on a Rogers 4350B substrate and impedance matched to 50 $\Omega$ through a three-stage quarter-wave transformer (Fig.~\ref{fig:Antenna}b(i)). The other side of the substrate is a partial ground metal plane of copper (Fig.~\ref{fig:Antenna}b(ii)). The patch radiates into a polystyrene-based dielectric cavity which is enclosed, on the other end, by a Rogers 4350B superstrate. The inner side of the superstrate is printed with a periodic array of $4 \times 4$ unit cells in copper to form a PRS (Fig.~\ref{fig:Antenna}b(iii)). Each unit cell of the PRS is a rectangular loop with a diagonal. The dimensions of this basic antenna structure (Fig.~\ref{fig:Antenna}) remain fixed across all candidate designs. Then, peripheral roughness is introduced to metallic edges along each dimension of the unit cell through 36 $0.5 \times 0.5$ mm$^2$ metal bricks (Fig.~\ref{fig:Antenna}b(iv)). 

The positions of the bricks along the peripheries of the rectangular loop become the DoFs for reducing the AR while enhancing the gain and bandwidth. The position of each brick is indicated in two-dimensional (2-D) Cartesian coordinates, with the origin assumed to be at the left lower corner of the unit cell. The antenna system with each unique unit cell design is simulated in CST Microwave Studio to obtain the electrical characteristics from 2 to 3 GHz. Since this is a 3-D antenna structure, there are approximately 6.4 million mesh cells for each design, with a simulation duration of each around 75 minutes, which is excessive. 
\section{GAN Architecture for CP FPC Design}
\label{sec:GAN Architecture for Antenna Design}
\begin{figure}[t]
\centering
\includegraphics[scale=0.6]{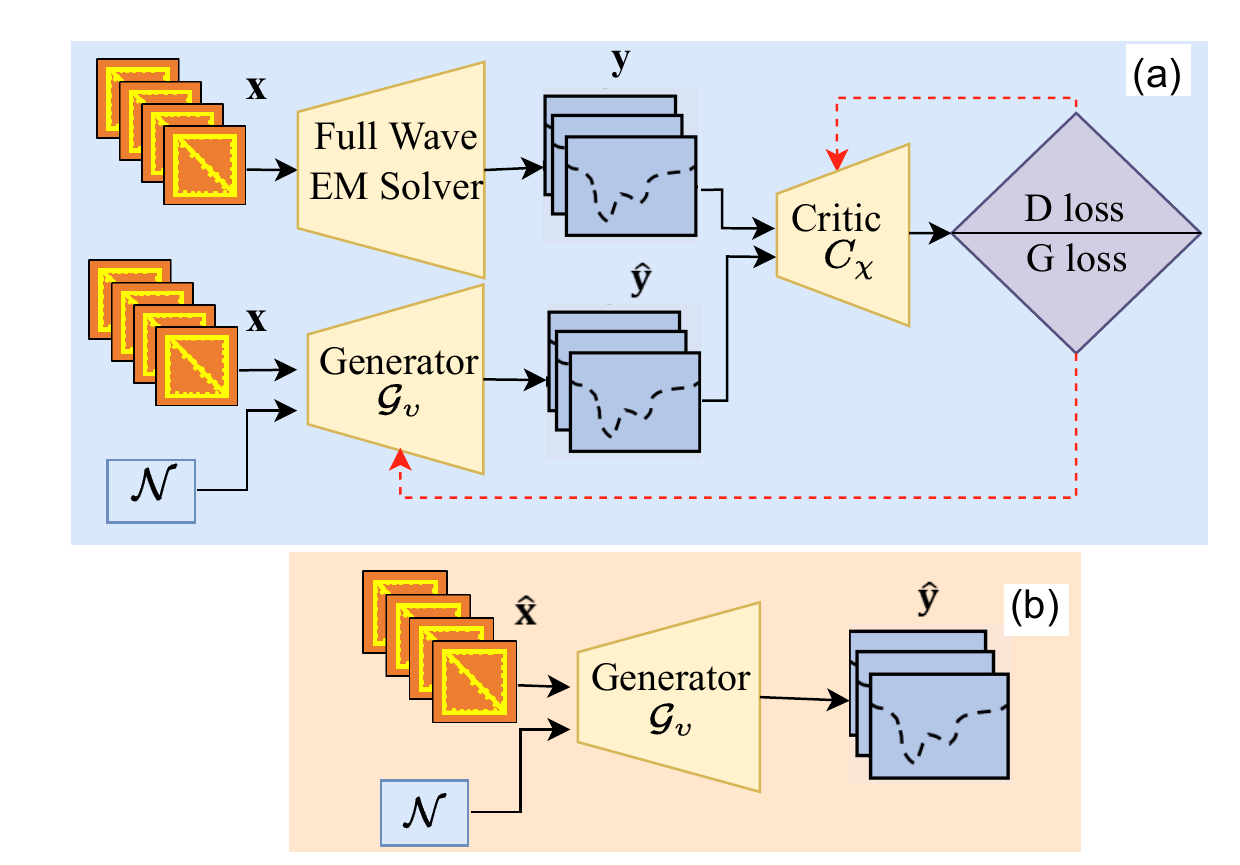}
\vspace{-3mm}
\caption{GAN architecture for generating surrogate models of antenna behavior for an input antenna design (a) training and (b) prediction stages.}
\label{fig:TrainingGAN}
\end{figure}

We propose to reduce the resource-expensive EM simulations by replacing them with a GAN at the prediction stage. This network comprises of a generator ($\mathcal{G}_{\upsilon}$) and a critic ($C_{\chi}$) (Fig.~\ref{fig:TrainingGAN}a). During the training phase, the 2-D position coordinates of 36 bricks in the unit cell in the PRS were reshaped to a single column vector, $\xb$, of size $[72 \times 1]$.
This was concatenated with a latent noise vector, $\mathcal{N}$, of $[100 \times 1]$ size, to prevent overfitting, and provided as input $\mathbf{z}=(\mathcal{N} \parallel \xb$), of $[172 \times 1]$ size to $\mathcal{G}_{\upsilon}$. The output of the generator, $\y$, were antenna performance metrics (as a function of frequency): AR, return loss, and gain, each of size $[101 \times 1]$ concatenated to form a single column vector. These were input to $C_{\chi}$ along with the real antenna characteristics, $\yb$, obtained from CST Microwave Studio for the same set of input antenna designs $\xb$. Both GAN networks compete adversarially to optimize the weights $\upsilon$ and $\chi$ of, respectively, $\mathcal{G}_{\upsilon}$ and $\mathcal{C}_{\chi}$ based on the value function 
\begin{small}
\begin{align}
\label{eqn:objectivefn}
\nonumber V(\mathcal{G}_{\upsilon},\mathcal{C}_{\chi})=
\mathbb{E}_{y \sim p_{y}(y)}[log(\mathcal{C}_{\chi}(y))]+\mathbb{E}_{\mathbf{z} \sim p_{\mathbf{z}}(\mathbf{z}})[\log(1-\mathcal{C}_{\chi}(\mathcal{G}_{\upsilon}(\mathbf{z})].
\end{align}
\end{small}
The training process involved iterative simultaneous stochastic gradient descent based on Adam optimization on batches of 16 samples of $\xb, \yb$ and $\hat{\yb}$. In each iteration, ${\upsilon}$ was updated while ${\chi}$ kept constant and vice versa. The weights were normalized during each update to prevent overfitting and weighted by a regularizer $\lambda=0.01$.
\begin{figure}[t]
\centering
\includegraphics[scale=0.33]{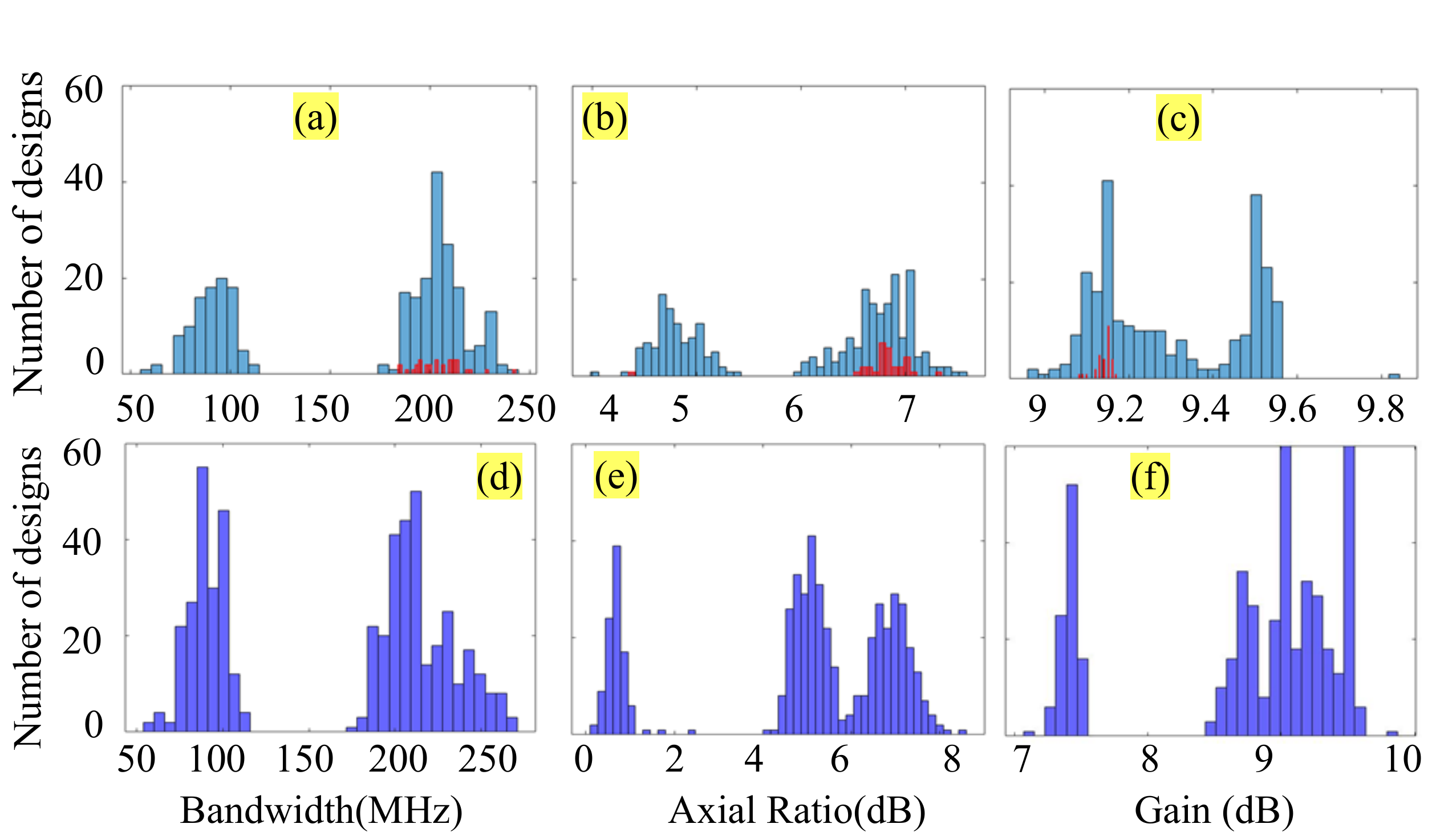}
\vspace{-3mm}
\caption{Histograms depicting variations of antenna characteristics for (a-c) 300 samples that were used to generate GAN and CST results; both training (blue) and test (red) sets are shown (d-f) 500 candidate antenna designs from which an optimized design was selected. }
\label{fig:Histogram}
\end{figure}
The total number of iterations was set to 10000. The learning rate of the stochastic gradient descent operation for both networks was $5\times10^{-4}$, and batch normalization was applied with a momentum of 0.8. 
The network $G$ had 128, 256, and 512 nodes in the first, second, and third layers, respectively, with \emph{Leaky ReLU} activation functions. The network $C$ had similar 512 and 256 nodes in two respective hidden layers. The output layer had one node with the $sigmoid$ function. 
\section{Numerical Experiments}
\label{sec:numexp}
\begin{figure}[t]
\centering
\includegraphics[scale=0.46]{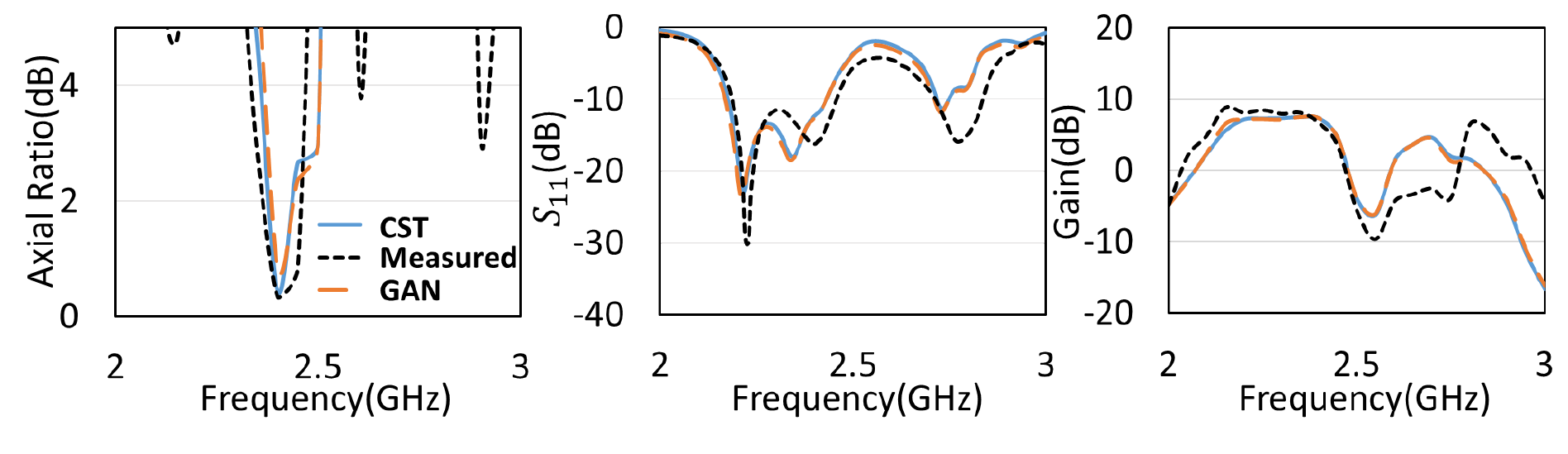}
\vspace{-7mm}
\caption{Antenna performance metrics for surrogate GAN model, CST, and fabricated device measurements.}
\label{fig:ExperimentalResult}
\end{figure}

We implemented the DL network with Keras 2.7 and Python on an Intel Core i7-10510U processor running at 1.80 GHz and NVIDIA GeForce MX250.
\emph{Validation of surrogate model via simulations:}
Consider the architecture in Fig.~\ref{fig:TrainingGAN}b where the input antenna designs (distinct from those used for training) correspond to the brick positions in the unit cell form $\mathbf{\hat{x}}$ and the output are the corresponding antenna characteristics $\y = \mathcal{G}(\mathbf{z})$. The (fake) output $\y$ of $\mathcal{G}$ was then compared with $\yb$ generated from the EM solver to monitor the training process. We used $90\%$ and $10\%$ of a total of 300 antenna designs-pattern pairs for training and validation, respectively, with 10-fold cross-validation. We compared the performance metrics as a function of frequency from 2 to 3 GHz. 
\begin{table}[t]
\centering
\caption{ NMSE of antenna characteristics  }
\label{method_Comparison}
\begin{tabular}{p{1cm}|p{2cm}|p{2cm}|p{2.0cm}}
\hline 
\noalign{\vskip 1pt}
Methods  & Gain & Axial Ratio & Return Loss\\[1pt]
\hline \hline
\noalign{\vskip 1pt}
     MLP& 0.3&0.09&0.27 \\
     CNN&0.28 &0.08&0.26\\
     GAN&0.23&0.05&0.2\\
     \hline 
\end{tabular}
\vspace{-4mm}
\end{table}
Training and test data in Fig.~\ref{fig:Histogram}a-c show that these metrics vary with the spatial distribution of the peripheral roughness features, and the data were not overfitted. Finally, comparisons (Table.~\ref{method_Comparison}) with a multi-layer perceptron (MLP) and a convolutional neural network (CNN) trained with the same data demonstrate that GAN has the lowest normalized mean square error (NMSE) between the real and fake antenna characteristics.\\
\emph{Fabrication and measurement results:}
\label{subsec:GAN Rsults}
The simple rectangular patch is essentially an LP narrowband (20 MHz bandwidth) and low gain (3.4 dBi) antenna. The PRS structure enhances the gain and bandwidth of the structure to 188.5 MHz, and 9.4 dBi gain (Fig.~\ref{fig:ExperimentalResult}) with a resonant frequency shifted from 2.4 GHz.
The unit cell design of a \emph{rectangular loop with a diagonal} without roughness transforms the patch’s LP signal to an elliptically polarised wave with an AR of 7.4 dB. The peripheral roughness features reduce the AR without accounting for DoF in the metal brick distribution. The trained generator is able to analyze the antenna characteristics of 500 such designs in 20 seconds (as against 625 hours with CST). From these 500 antenna characteristics (shown in Fig.~\ref{fig:Histogram}d-f), we chose the design with a wide bandwidth of 269 MHz and an AR of 0.4 dB. We validated this with CST and fabricated the corresponding antenna (Fig.~\ref{fig:Antenna}c). The measurements of gain, return loss, and the axial ratio is carried out with a vector network analyzer N9926A  and a linearly polarized reference horn antenna (HF907) with known gain characteristics. The measurements from the actual antenna show very good agreement with GAN and CST results (Fig.~\ref{fig:ExperimentalResult}). 
\begin{table}[htbp]
\centering
\begin{threeparttable}
\caption{Comparison with competing antenna structures}
\label{table:Comparison_antenna}
\begin{tabular}{p{2cm}|p{1.5cm}|p{1cm}|p{1cm}|p{1cm}}
\hline 
\noalign{\vskip 1pt}
    Antenna& ZBW (MHz)&5dB BW (MHz)& Axial Ratio(dB)& Gain(dBi) \\
\hline \hline
\noalign{\vskip 1pt}
     Simple patch & 20 &-\tnote{1} & 40 & 3.4\\
     FPC with PRS & 88.5 &-\tnote{2}& 7.6 & 9.4\\
     GAN-based FPC,rough PRS  & 269 &100 & 0.4 & 7.5\\ 
     \hline 
\end{tabular}
\begin{tablenotes}
\item[1] Simple patch is linearly polarized and hence does not have a 5dB AR bandwidth. 
\item[2] FPC with PRS is elliptically polarized with the lowest AR of 7.6dB.
\end{tablenotes}
\end{threeparttable}
\end{table}
Quantitative comparisons (Table.\ref{table:Comparison_antenna}) with FPC and the simple patch further demonstrate the performance enhancements with a GAN-aided design. We consider four antenna characteristics - the gain at the resonant frequency at the antenna boresight, the return loss bandwidth (ZBW), the 5dB AR bandwidth, and the corresponding optimum AR. We observe that the proposed antenna has a wider bandwidth, lower AR, and higher gain than the simple patch. The proposed antenna also has a wider bandwidth and lower AR with respect to the FPC with smooth PRS, though the gain is slightly lower.
\section{Summary}
\label{sec:Conclusion}
Peripheral roughness along the edges of the unit cell of the PRS of a CP-FPC offers several DoFs for improving antenna performance. We proposed a versatile GAN-based FPC design strategy where we train a GAN to serve as a surrogate model using input-output pairs of antenna designs and their corresponding patterns obtained from the solver. The proposed design strategy enables rapid evaluation of a large number of candidate designs. Our GAN-optimized unit cell yielded AR, gain, and bandwidth of 0.4 dB, 7.5 dBi, and 269 MHz, respectively, thereby considerably improving the performance of the original FPC structure. Fabrication and experimental validations supported the GAN results. 
The design files, GAN codes, and supplementary document( with the description of antenna design parameter and GAN network hyper-parameters ) are available at \url{https://essrg.iiitd.edu.in/?page_id=4355}.

\bibliographystyle{ieeetr}
\bibliography{main}

\end{document}